\documentclass{jssc}

\usepackage{amsfonts}
\usepackage{epsfig}
\usepackage{amsmath}
\usepackage{graphicx}
\usepackage{amsmath}
\usepackage{enumerate}
\usepackage{amscd}
\usepackage{stmaryrd}
\usepackage{algorithm,float}
\usepackage{arydshln}
\usepackage{algorithmic}
\usepackage{youngtab}
\usepackage{young}
\usepackage{geometry}
\usepackage{bbm}
\usepackage{leftidx}
\usepackage{multirow}
\usepackage{url}

\usepackage[bookmarks,
            colorlinks,
            linkcolor=red,
            anchorcolor=red,
            citecolor=red,
            ]{hyperref}

\def\textsubscript#1%
{$_{\text{#1}}$}

\def\cdd{\mbox{\boldmath$\cdot$}~}

\def\ee{{\rm e}}

\input{amssym.def}
\include{graphix}

\def\be{\begin{equation}}
\def\ee{\end{equation}}
\def\bd{\begin{description}}
\def\ed{\end{description}}
\def\bi{\begin{itemize}}
\def\ei{\end{itemize}}
\def\ba{\begin{array}}
\def\ea{\end{array}}
\def\bu{\begin{enumerate}}
\def\eu{\end{enumerate}}

\def\ds{\displaystyle}

\makeatletter
\newenvironment{breakablealgorithm}
  {
   \begin{center}
     \refstepcounter{algorithm}
     \hrule height.8pt depth0pt \kern2pt
     \renewcommand{\caption}[2][\relax]{
       {\raggedright\textbf{\ALG@name~\thealgorithm} ##2\par}%
       \ifx\relax##1\relax 
         \addcontentsline{loa}{algorithm}{\protect\numberline{\thealgorithm}##2}%
       \else 
         \addcontentsline{loa}{algorithm}{\protect\numberline{\thealgorithm}##1}%
       \fi
       \kern2pt\hrule\kern2pt
     }
  }{
     \kern2pt\hrule\relax
   \end{center}
  }
\makeatother

\makeatletter
\def\@oddfoot{\hfill}
\newcount\shumeicount
\def\setshumei#1#2#3{%
  \shumeicount=\count0
  \def\@oddhead{%
    \raise-5pt\hbox to0pt{\vrule width\hsize height 0pt depth 0.4pt\hss}\relax
    \ifnum \shumeicount=\count0
      \raise-7pt\hbox to0pt{\vrule width\hsize height 0pt depth 0.4pt\hss}\relax
      #1
    \else
      \ifodd\count0
        #2
      \else
        #3
       \fi
     \fi
  }%
}
\makeatother
\makeatletter
\def\@oddfoot{\hfill}
\newcount\shujiaocount
\def\setshujiao{%
  \shujiaocount=\count0
  \def\@oddfoot{%
      \ifodd\count0
      \else
      \fi
  }%
}
\makeatother
\def\biaoti#1#2#3#4{{
  \vspace*{0.3cm}
  \begin{flushleft} \Large\bf #1\end{flushleft}
  \vspace*{-0.2cm}
      \begin{flushleft}
      \bf #2
      \end{flushleft}
      \footnotetext{\hspace{-6mm} #3\\ #4}}}

\def\dshm#1#2#3#4
{\setshumei{
{\thepage--\pageref{LastPage}}\hfill}
            {\hfill {\small #3}\hfill\hbox to0pt{\hss\thepage}}
            {\hbox to0pt{\thepage\hss}\hfill {\small #4}\hfill
            }
            \setshujiao}
\def\drd#1#2
{{\vskip 1cm\small \begin{flushleft}
 #1 \\
 #2\\
\copyright 
\end{flushleft}}}

\def\epsilon{\varepsilon}

\begin{document}

\biaoti{A New Quantum Algorithm for the Random Subset Sum Problem}
{Yang Li \cdd
Hongbo Li }
{Address: Academy of Mathematics and Systems Science,
University of Chinese Academy of Sciences,
Chinese Academy of Sciences, Beijing 100190, China\\
    Email: liyang815@mails.ucas.edu.cn  $\cdot$ hli@mmrc.iss.ac.cn} 
{\\
{}}



\Abstract{Solving random subset sum instances
plays an important role in constructing cryptographic systems. For the random subset sum problem,
in 2013 Bernstein et al. proposed a quantum
algorithm with heuristic time complexity $\widetilde{O}(2^{0.241n})$,
where the ``$\widetilde{O}$" symbol is used to omit
poly($\log n$) factors.
In 2018, Helm and May proposed another quantum algorithm that reduces the heuristic
time and memory complexity to $\widetilde{O}(2^{0.226n})$. In this paper, a new quantum
algorithm is proposed, with heuristic time and memory
complexity $\widetilde{O}(2^{0.209n})$.}      




\section{Introduction}\label{sec:1}

The subset sum problem ({\bf SSP})
is a fundamental problem in theoretical computer science,
and is one of the most famous NP-hard problems\cite{gj79}. Due to its difficulty, the subset
sum problem is polular in designing cryptosystems\cite{cr84,fmv16,in96,lps10,mh78}.

Given data $(a_1,a_2,\dots,a_n,s)\in(\mathbb{Z}_{2^{n}})^{n+1}$, called an
{\it instance}, there exist two forms of {\bf SSP}. The first is the decision
{\bf SSP}: decide whether there exists a subset of $\{a_1,a_2,\dots,a_n\}$ summing up to
$s$; or in vector form, decide whether there exists a vector $\mathbf{e}\in\{0,1\}^n$ such
that
\begin{equation}
\langle\mathbf{a},\mathbf{e}\rangle\equiv s \mod 2^{n}.
\label{def:ssp}
\end{equation}
The second is the
computational {\bf SSP}: find a subset of $\{a_1,a_2,\dots,a_n\}$
summing up to $s$.
The decision {\bf SSP} is NP-complete. Given
access to an oracle that solves the decision {\bf SSP}, the computational {\bf SSP} can be
solved with $n$ calls to this oracle.

If the data $(a_1,a_2,\dots,a_n,s)\in(\mathbb{Z}_{2^{n}})^{n+1}$ is given randomly,
the subset sum problem becomes the {\it random subset sum problem} ({\bf RSSP}).

\begin{definition}
Let $\mathbf{a}\in(\mathbb{Z}_{2^{n}})^{n}$ be chosen at random uniformly.
For a random $\mathbf{e}\in\{0,1\}^n$ with $|\mathbf{e}|=\frac{n}{2}$, let
$s\equiv\langle\mathbf{a},\mathbf{e}\rangle \mod  2^{n}$, where $|\mathbf{e}|$ stands for
the {\it Hamming weight} of $\mathbf{e}$. Then $(\mathbf{a},s)\in(\mathbb{Z}_{2^{n}})^{n+1}$
is called a {\it random subset sum instance}.
The {\it density} of the random subset sum instance is
\be
d:=\frac{n}{\log (\max_i a_i)}.
\ee

Every $\mathbf{f}\in\{0,1\}^n$ satisfying
$\langle\mathbf{a},\mathbf{f}\rangle\equiv s\mod  2^{n}$ is called a {\it solution} to the
random subset sum instance.
The {\it random subset sum problem} ({\bf RSSP}) refers to the problem of
finding a solution to a
random subset sum instance.
\end{definition}

\vskip .2cm
{\bf {\bf SSP} and {\bf RSSP} algorithms running on classical computer.}
\vskip .2cm

For {\bf SSP},
enumerating all possible $\mathbf{e}\in\{0,1\}^n$ and checking whether
$\langle\mathbf{a},\mathbf{e}\rangle\equiv s\mod  2^{n}$ can solve this problem
in time $\widetilde{O}(2^{n})$. In 1974, Horowitz and Sahni (HS)\cite{hs74} introduced
a Meet-in-the-Middle algorithm with time and space complexity
$\widetilde{O}(2^{\frac{n}{2}})$.
In the HS algorithm, enumerating all $\mathbf{e}_1\in\{0,1\}^{\frac{n}{2}}\times
0^{\frac{n}{2}}$, $\mathbf{e}_2\in 0^{\frac{n}{2}}\times\{0,1\}^{\frac{n}{2}}$
and setting up two ordered lists $L_1,L_2$, that sorted by $\langle\mathbf{a},
\mathbf{e}_1\rangle,s-\langle\mathbf{a},\mathbf{e}_2\rangle$ accordingly.
Then, for each $(\mathbf{e}_1,\langle\mathbf{a},\mathbf{e}_1\rangle)\in L_1$,
looking for $(\mathbf{e}_2,s-\langle\mathbf{a},\mathbf{e}_2\rangle)\in L_2$
that satisfies $s-\langle\mathbf{a},\mathbf{e}_2\rangle\equiv\langle\mathbf{a},
\mathbf{e}_1\rangle\mod  2^{n}$ by binary search.
If there is a collision $\langle\mathbf{a},\mathbf{e}_1\rangle\equiv s
-\langle\mathbf{a},\mathbf{e}_2\rangle\mod  2^{n}$, then
$\mathbf{e}_1+\mathbf{e}_2$ is a solution.

In 1981, Schroeppel and Shamir (SS) \cite{ss81} improved this to time complexity
$\widetilde{O}(2^{\frac{n}{2}})$ with only space complexity
$\widetilde{O}(2^{\frac{n}{4}})$.  These algorithms are still the fastest known
for solving general instances of subset sum.

For {\bf RSSP}, Brickell \cite{b84}, Lagarias and Odlyzko
\cite{lo85} showed that random subset sum instances can be solved with density $d<0.64$,
by giving an oracle solving the shortest vector problem $(SVP)$ in lattices.

In 1991 this bound was improved by Coster et al. \cite{clos91} and Joux, Stern \cite{js91} to $d<0.94$. Note that this transformation does not rule out the hardness of subset sum problem in the low-density regime, since solving $SVP$ is known to be NP-hard \cite{a98}. In the high-density regime with $d=\Omega(\frac{1}{log\;n})$ dynamic programming solves subset sum problem efficiently \cite{gm91}.

However, for the case $d\approx 1$ only exponential time algorithms are known. In a breakthrough paper, Howgrave-Graham and Joux (HGJ) \cite{hgj10} at Eurocrypt 2010 showed that random subset sum instances can be solved in time $\widetilde{O}(2^{0.337n})$. The main technique used is called representation technique. Recall that the HS algorithm splits $\mathbf{e}$ as $\mathbf{e}_1\in\{0,1\}^{\frac{n}{2}}\times 0^{\frac{n}{2}}$ and $\mathbf{e}_2\in 0^{\frac{n}{2}}\times\{0,1\}^{\frac{n}{2}}$. The main idea of HGJ is to represent $\mathbf{e}$ in a different, ambiguous way as a $4$-sum $\mathbf{e}_1 +\mathbf{e}_2 +\mathbf{e}_3 +\mathbf{e}_4$ with $\mathbf{e}_{i}\in\{0,1\}^n,|\mathbf{e}_i|=\frac{n}{8}, 1\le i\le 4$. As a consequence, the HGJ technique is called in the literature {\it representation technique}. The HGJ algorithm first constructs $4$ lists of candidates $\mathbf{c}_i\in\{0,1\}^n$ for $\mathbf{e}_i$ by enumerating all candidates $\mathbf{c}_i\in\{0,1\}^n$ with $|\mathbf{c}_i|=\frac{n}{8}$. It then computes $2$-sums $\mathbf{c}_1+\mathbf{c}_2,\mathbf{c}_3+\mathbf{c}_4\in\{0,1,2\}^n$ and filters out all sums that contain $2$-entries. To control the list sizes (which in turn
determine the run time), some constraints are introduced. At the same time, these constraints reduce the number of representations. The key observation is that finding only one representation of $\mathbf{e}$ is sufficient to solve the random subset sum problem. Therefore, the parameters in HGJ need to be optimized based on this objective.

At Eurocrypt 2011, Becker, Coron and Joux (BCJ) \cite{bcj11} proposed a modification to the HGJ algorithm with heuristic run time $\widetilde{O}(2^{0.291n})$. The core idea of the BCJ algorithm is to represent $\mathbf{e}$ as an $8$-sum $\mathbf{e}_1+\dots+\mathbf{e}_8$ with $\mathbf{e}_i\in\{-1,0,1\}^n, 1\le i\le 8$. The BCJ algorithm as well as the HGJ algorithm proceed in a divide-and-conquer fashion. The BCJ algorithm first uses enumeration to construct $8$ lists of candidates $\mathbf{c}_i\in\{-1,0,1\}^n$ with a certain pre-defined (optimized) number of $-1$'s, $0$'s and $1$'s. It then computes $2$-sums $\mathbf{c}_1+\mathbf{c}_2,\dots,\mathbf{c}_7+\mathbf{c}_8\in\{-2,-1,0,1,2\}^n$ and filters out all sums that contain $\pm2$-entries, and in addition filters out among
all remaining vectors those that do not possess another pre-defined (optimized) number of $-1$'s, $0$'s and $1$'s. As above, the parameters in BCJ need to be optimized to make sure that one representation of $\mathbf{e}$ can be found.

In 2019, Esser and May (EM) \cite{em19} proposed a new heuristic algorithm based on representation and sampling technique with run time $\widetilde{O}(2^{0.255n})$. While the initial lists in HGJ and BCJ are constructed by enumeration, the initial lists in EM are constructed by sampling from a Bernoulli distribution. Sampling technique introduces variance that increases the amount of representations and brings more optimization flexibility. Note that all lists in EM form a tree. A remarkable property is that the complexity of the EM algorithm improves with increasing tree depth.

\vskip .2cm
{\bf {\bf SSP} and {\bf RSSP}  algorithms running on quantum computer.}
\vskip .2cm

In 2013, Bernstein, Jeffery, Lange and Meurer \cite{bjlm13} constructed quantum subset sum algorithms, inspired by the HS algorithm, the SS algorithm and the HGJ algorithm. In detail, Bernstein et al. showed that the quantum HS algorithm achieve run time $\widetilde{O}(2^{n/3})$. Moreover, a first quantum version of the SS algorithm with Grover search \cite{g96} runs in time $\widetilde{O}(2^{3n/8})$ using only space $\widetilde{O}(2^{n/8})$. A second quantum version of the SS algorithm using quantum walks \cite{aakv01,a07} achieves time $\widetilde{O}(2^{0.3n})$. Eventually, Bernstein et al. used the quantum walk framework of Magniez et al. \cite{mnrs11} to achieve a quantum version of the HGJ algorithm with time and space complexity $\widetilde{O}(2^{0.241n})$. In 2018, Helm and May \cite{hm18} achieve a quantum version of the BCJ algorithm with time and space complexity $\widetilde{O}(2^{0.226n})$, which is the best known quantum random subset sum algorithm.

Quantum algorithms based on the quantum walk framework are designed in the following three steps:

(1). Start with a classic algorithm.

(2). Generalize to a lower-probability algorithm and build a data structure that expresses the entire computation of the lower-probability algorithm.

(3). Apply a quantum walk.

The key point of quantum HGJ and quantum BCJ algorithms is that we no longer enumerate the initial lists, but only start with random subsets of the initial lists with some fixed size that has to be optimized. On the one hand, subsets of the leaves lists yields small list sizes, which speeds up the construction of lists. On the other hand, subsets of the leaves lists reduces the probability that the corresponding classical algorithms succeed. The quantum algorithms achieve the acceleration of the corresponding classical algorithms because quantum walks amplify the probability of success.

\vskip .2cm
{\bf Contribution of this paper.}
\vskip .2cm

We propose a new quantum algorithm with running time down to $\widetilde{O}(2^{0.209n})$. Our algorithm is actually a quantum version of the EM algorithm. Note that the initial
lists in EM are constructed by sampling from a Bernoulli distribution. Recall that the initial lists in HGJ are constructed by enumerating all candidates $\mathbf{c}_i\in\{0,1\}^n$ with $|\mathbf{c}_i|=\frac{n}{8}$. Now consider how the initial lists are built in  EM. All we know is that the elements of the initial lists belongs to $\{0,1\}^n$. The existence of randomness prevents us from using quantum walks directly. One simple way to solve this problem is firstly sampling to give us the initial lists. Next, carry out quantum walks. Moreover, we need to define an appropriate quantum walk for the EM algorithm within the framework of Magniez et al. \cite{mnrs11}.

Note that, whereas the complexity of the EM algorithm improves with increasing tree depth, our quantum algorithm is optimal when the search depth is $4$.

The paper is organized as follows. In Section \ref{sec:2} we outline the quantum walk
technology and the EM classical algorithm. In Section \ref{sec:3} we firstly describe
the connection between random subset sum problem and graph search problem. Then we define
an appropriate data structure and give our quantum algorithm.

\section{Preliminaries}\label{sec:2}

By $H(\cdot)$ we refer to the binary entropy function, which is defined on input $0\le \alpha\le 1$ as $H(\alpha) := -\alpha \log \alpha-(1-\alpha)\; log\;(1-\alpha)$, where $log$ is the logarithmic function with base $2$ and we use the convention $0\; log\; 0 := 0$. We approximate binomial coefficients by the entropy function, derived from Stirling’s formula ${n \choose m}=\widetilde{\Theta}(2^{nH(m/n)})$.

Let $X \sim \mathcal{D}$ be a discrete random variable following the distribution $\mathcal{D}$, which is defined on a finite alphabet $\Lambda$. For $x\in \Lambda$ let $p_X(x) := Pr [X = x]$. We define the entropy of a random variable or equivalently its distribution as
\begin{displaymath}
H(X)=H(\mathcal{D}):=-\sum_{x\in \Lambda}p_X(x)\log p_X(x).
 \end{displaymath}

For $0\le \alpha \le 1$ we refer by $\mathcal{B}(\alpha)$ to the Bernoulli distribution with parameter $\alpha$, that is for $X \sim \mathcal{B}(\alpha)$ we have $Pr [X = 1] = \alpha$ and $Pr [X = 0] = 1-\alpha$. The sum of $m$ iid $\mathcal{B}(\alpha)$-distributed random variables is binomially distributed with parameters $m$ and $\alpha$, which we denote by $Bin_{m,\alpha}$.
Let $\mathbf{x} \sim Bin_{m,\alpha}^n$ denote a vector of $n$ iid random variables, thus $\mathbf{x}=(x_1,\dots,x_n)$ with $x_i \sim Bin_{m,\alpha}$
and therefore $x_i\in \{0,\dots,m\}$. Note that the entropy of such a vector is $H(\mathbf{x}) = H(Bin_{m,\alpha}^n)= H(Bin_{m,\alpha})n$.

\subsection{Quantum walks}\label{sec:2.1}

\begin{problem}[Graph Search Problem]\label{pro:2.1}
Given a graph $G=(V,E)$ and a set of vertices $M\subset V$, called the set of all marked vertices, find a marked vertex $u\in M$.
\end{problem}

The graph search problem can be solved using a quantum walk on graph $G$. A
state of the walk will correspond to a vertex $u\in V$, and a data structure $d(u)$ associated to each state $u$ will help us to decide whether $u$ is marked. Three types of cost are associated with $d(u)$. The setup cost $T_s$ is the cost to set up the data structure $d(u)$ for a given vertex $u\in V$. The update cost $T_u$ is the cost to update the data structure, i.e., the cost needed to convert $d(u)$ into $d(v)$ for two given connected vertices $u,v\in V$. The checking cost $T_c$ is the cost of checking with high probability whether $u$ is marked, given $u\in V$ and $d(u)$.

Several quantum walks algorithms have been proposed by many authors, notably Ambainis \cite{a07}, Szegedy \cite{s04}, and Magniez et al. \cite{mnrs11}. A survey of these results can be found in \cite{s08,sll19}. The following theorem is important and useful.

\begin{theorem}[Magniez et al. \cite{mnrs11}] \label{th:2.2}
Let $G = (V,E)$ be a regular graph with spectral gap $\delta$, and let $\epsilon > 0$ be a lower bound on the probability that a vertex chosen randomly of $G$ is marked.
Let $T_s,T_u,T_c$ be the setup, update and checking cost.
Then there exists a quantum algorithm that with high probability finds a marked vertex with cost

\begin{equation}\label{E:1}
T=T_s+\frac{1}{\sqrt{\epsilon}}(\frac{1}{\sqrt{\delta}}T_u+T_c).
\end{equation}

\end{theorem}

Note that $T$ is time (memory) complexity of this quantum algorithm if $T_s,T_u,T_c$ are measured in
time (memory).

\begin{definition}[Johnson Graph]\label{def:2.3}
Given a set $L$ with $|L|=N$, the Johnson graph $J=(N,r)$ is an undirected graph whose vertices are the subsets of $L$ containing $r$ elements, where $0\le r\le N$. An edge between two vertices $S$ and $S^{'}$ exists iff $|S\cap S^{'}| = r - 1$. That is, two vertices are adjacent iff they differ in only one element that belongs to $L$.
\end{definition}

\begin{definition}[Cartesian Product of Graphs]\label{def:2.4}
Let $G_1=(V_1,E_1),G_2=(V_2,E_2)$ be undirected graphs.The Cartesian product $G_1\times G_2=(V,E)$ is defined via

$V=V_1\times V_2=\{v_1v_2|v_1\in V_1,v_2\in V_2\}$ and

$E=\{(u_1u_2,v_1v_2)|(u_1=v_1\land (u_2,v_2)\in E_2)\lor ((u_1,v_1)\in E_1\land u_2=v_2) \}$

\end{definition}

For Johnson graphs it is well-known that $\delta(J(N,r))=\Omega(1/r)$. The following lemma gives us the spectral gap of the cartesian product of Johnson graphs.

\begin{lemma}[Kachigar, Tillich\cite{kt17}]\label{le:2.5}
Let $J (N, r)$ be a Johnson graph, and let $J^m (N, r) :=\times^{m}_{1}J (n, r)$. Then $\delta(J^m (N, r))\ge \frac{1}{m}\delta(J(N,r))$.
\end{lemma}

\subsection{The EM Classical Algorithm}\label{sec:2.2}

Denote by $EM^{(d)}$ the EM algorithm with tree depth $d$. Whereas run time of the EM algorithm decreases with increasing tree depth $d$, our quantum algorithm is optimal at the tree depth $4$. Thus, we only describe $EM^{(4)}$.

Let $(\mathbf{a},s)\in(\mathbb{Z}_{2^{n}})^{n+1}$ be a subset sum instance
with a solution $\mathbf{e}\in\{0,1\}^n$ with $|\mathbf{e}|=\frac{n}{2}$. That is, $\langle\mathbf{a},\mathbf{e}\rangle\equiv s\mod  2^{n}$.

The basic idea of representation is to represent the solution $\mathbf{e}$ as a sum $\mathbf{e}_1+\mathbf{e}_2$ where $\mathbf{e}_1$ and $\mathbf{e}_2\in\{0,1\}^n$. If two lists $L_1$ and $L_2$ whose elements are candidates for $\mathbf{e}_1,\mathbf{e}_2$ accordingly can effectively (respectively $\widetilde{O}(L_1),\widetilde{O}(L_2)$) be constructed, then using the list join operator above, we can effectively get the solution $\mathbf{e}$ via the joined list $L_0=\{\mathbf{e}=\mathbf{e}_1+\mathbf{e}_2|(\mathbf{e}_1,\mathbf{e}_2)\in L_1 \times L_2\land \langle\mathbf{a},\mathbf{e}_1\rangle+\langle\mathbf{a},\mathbf{e}_2\rangle\equiv s\mod  2^{n}\}$. Note that sorting $L_1$ and searching are performed with respect to $\langle\mathbf{a},\mathbf{e}_1\rangle$ and $\langle\mathbf{a},\mathbf{e}_2\rangle$, where $\mathbf{e}_1\in L_1,\mathbf{e}_2\in L_2$. Generally speaking, this effectiveness of constructing base lists $L_1,L_2$ is not guaranteed. Thus, splitting the solution $\mathbf{e}$ several times leads to improve the complexity of the EM algorithm. That is, represent the solution $\mathbf{e}$ as a sum $\sum_{j} \mathbf{e}_{j}$ for some $j$ and construct lists $L_j$ whose elements are candidates for $\mathbf{e}_j$ accordingly.

\begin{definition}[The Level-$i$ Representation]\label{def:2.6}
$(\mathbf{e}^{(i)}_1,\mathbf{e}^{(i)}_2,\cdots,\mathbf{e}^{(i)}_{2^{4-i}})$ is called a level-$i$ representation if every $\mathbf{e}^{(i)}_j\in\{0,1\}^n,1\le j\le 2^{4-i}$ and $\sum_j \mathbf{e}^{(i)}_j=\mathbf{e}$.
\end{definition}

In $EM^{(4)}$, the solution $\mathbf{e}$ is represented as a sum $\sum_{j} \mathbf{e}^{(i)}_{j}$ with $\mathbf{e}^{(i)}_{j}\in\{0,1\}^n$, where $1\le i\le 4$, $1\le j\le 2^{4-i}$. Construct lists $L^{(i)}_{j}$ whose elements are candidates for $\mathbf{e}^{(i)}_{j}$ accordingly. Then all lists form a tree. The tree structure of $EM^{(4)}$ is shown in Figure $1$. Define the join operator for $k\in\mathbb{N},0\le k\le n$ and $s^{(i)}_j\in \mathbb{Z}_{2^{k}}$ as
$L^{(i)}_{j}=L^{(i-1)}_{2j-1} \bowtie_{(k,s^{(i)}_j)} L^{(i-1)}_{2j}:=\{\mathbf{x}_1+\mathbf{x}_2|(\mathbf{x}_1,\mathbf{x}_2)\in L^{(i-1)}_{2j-1} \times L^{(i-1)}_{2j}\land\langle \mathbf{a},\mathbf{x}_1+\mathbf{x}_2\rangle\equiv s^{(i)}_j\mod  2^{k}\}$, where $1\le i\le 4$, $1\le j\le 2^{4-i}$. Also write $L^{(i)}_{j}=L^{(i-1)}_{2j-1} \bowtie_{k} L^{(i-1)}_{2j}$.

Consider the join operator. On the one hand, these constraints $\langle \mathbf{a},\mathbf{x}_1+\mathbf{x}_2\rangle\equiv s^{(i)}_j\mod  2^{k}$ reduce the search space $|L^{(i)}_{j}|$. On the other hand, these constraints reduce the number of level-$i$ representations. The crucial observation is that it is sufficient to construct a single level-$4$ representation $(\mathbf{e}^{(4)}_1)$ of $\mathbf{e}$ in list $L^{(4)}_{1}$ for solving the original problem.

\begin{center}
  \centerline
  {\includegraphics[scale=0.35]{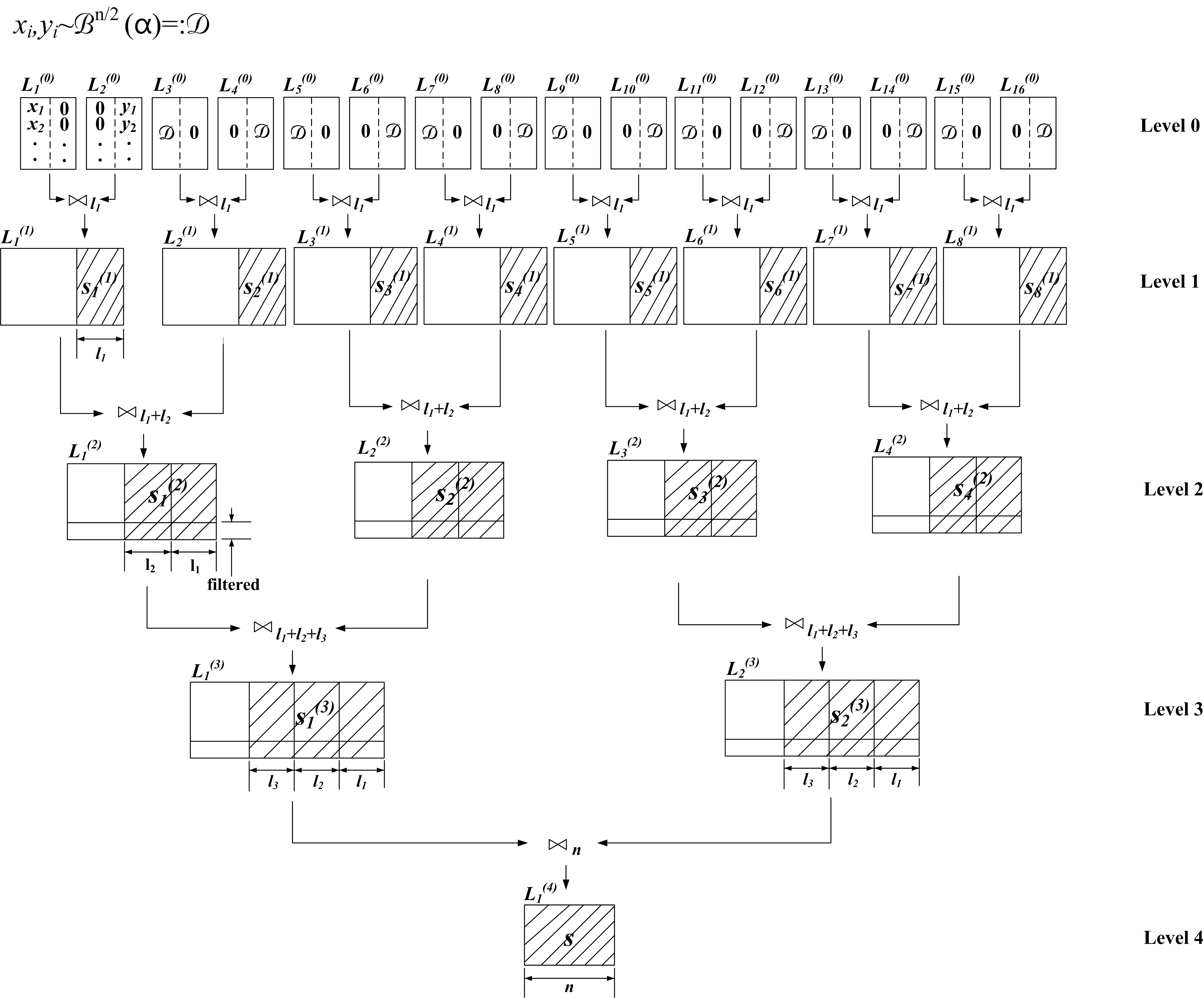}}\vskip3mm
\centering{\small {\bf Figure 1}\ \ Tree structure of $EM^{(4)}$. The portion covered by the slash represents the part of the lists that meet the constraints accordingly. \label{fig1}}
\end{center}

{\bf The Process of $EM^{(4)}$}

To construct on expectation a single level-$4$ representation, initially construct the level $0$ lists $L^{(0)}_{2j-1}$, $L^{(0)}_{2j}$, where $1\le j\le 8$. For each j, sample iid vectors $(\mathbf{x},0^{n/2})\in L^{(0)}_{2j-1}$, and $(0^{n/2},\mathbf{y})\in L^{(0)}_{2j}$, where $\mathbf{x},\mathbf{y}\sim\mathcal{B}^{\frac{n}{2}}(\alpha)\in\{0,1\}^{n/2}$.
Then, construct the level $1$ lists $L^{(1)}_{j}=L^{(0)}_{2j-1} \bowtie_{l_1} L^{(0)}_{2j}$, where $1\le j\le 8$. Choose random $s^{(1)}_{1},s^{(1)}_{2},\dots,s^{(1)}_{7}\in \mathbb{Z}_{2^{l_1}}$, and let $s^{(1)}_{8}\equiv s-\sum^7_{j=1}s^{(1)}_{j}\mod 2^{l_1}$. By the definition of the join operator, on level $1$ we get only those candidates $\mathbf{c}^{(1)}_{j}\in L^{(1)}_{j}$ satisfying $\langle \mathbf{a},\mathbf{c}^{(1)}_{j}\rangle\equiv s^{(1)}_{j}\mod  2^{l_1}$ for some $0\le l_1\le n,1\le j\le 8$. Note that all level-$1$ candidates $\mathbf{c}^{(1)}_{j}$ are vectors from $\{0,1\}^n$.

Similarly, construct the level $2$ lists $L^{(2)}_{j}=L^{(1)}_{2j-1} \bowtie_{l_1+l_2} L^{(1)}_{2j}$, where $1\le j\le 4$. Note that $s^{(2)}_{j}$ be chosen randomly on $\mathbb{Z}_{2^{l_1+l_2}}$ satisfying $s^{(2)}_{j}\equiv s^{(1)}_{2j-1}+s^{(1)}_{2j}\mod 2^{l_1},1\le j\le 3$ and $s^{(2)}_{4}\equiv s-s^{(2)}_{1}-s^{(2)}_{2}-s^{(2)}_{3}\mod 2^{l_1+l_2}$. Then, construct the level $3$ lists $L^{(3)}_{j}=L^{(2)}_{2j-1} \bowtie_{l_1+l_2+l_3} L^{(2)}_{2j}$, where $1\le j\le 2$.
Note that $s^{(3)}_{1}$ be chosen randomly on $\mathbb{Z}_{2^{l_1+l_2+l_3}}$ satisfying $s^{(3)}_{1}\equiv s^{(2)}_{1}+s^{(2)}_{2}\mod 2^{l_1+l_2}$, and $s^{(3)}_{2}\equiv s-s^{(3)}_{1}\mod 2^{l_1+l_2+l_3}$.
Finally, construct $L^{(4)}_{1}=L^{(3)}_{1} \bowtie_{n} L^{(3)}_{2}$ by setting $s^{(4)}_{1}=s$. If $\exists\mathbf{c}^{(4)}_{1}\in L^{(4)}_{1}$ satisfying $|\mathbf{c}^{(4)}_{1}|=\frac{n}{2}$, then $\mathbf{c}^{(4)}_{1}$ is a solution of the original random subset sum instance.

Note that any non-binary $\mathbf{c}^{(i)}_{j}\in L^{(i)}_{j}$ cannot be part of a valid representation of $\mathbf{e}$,
and may safely be filtered out. Therefore, after constructing each $L^{(i)}_{j}$, immediately eliminate all non-binary vectors.

A pseudocode description of the $EM^{(4)}$ algorithm is given by Algorithm \ref{al1}.

\begin{breakablealgorithm}\label{al1}
\caption{$EM^{(4)}$}
\begin{algorithmic}[1]
{\bf Input:} subset sum instance $(\mathbf{a},s)\in(\mathbb{Z}_{2^{n}})^{n+1}$;

parameters $\alpha\in(0,1)$, and
$l_1,l_2,l_3\in\mathbb{N}$ with $l_1+l_2+l_3\le n$.

{\bf Output:} solution $\mathbf{e}\in\{0,1\}^{n}$ with $|\mathbf{e}|=\frac{n}{2}$, or
$\perp$ if no solution is found.

\STATE
Sample all level-$0$ lists $L^{(0)}_{i}$ for $i=1,\dots,16$ as following:

(1.1) Initially set
$L^{(0)}_{i}=\varnothing, i=1,\dots,16$.

(1.2)
for $i=1$ to $8$, do the following:

\hskip .5cm
(1.2.1)
Repeat $L^{(0)}_{2i-1}\gets L^{(0)}_{2i-1}\cup\{(\mathbf{x},0^{n/2})\in\{0,1\}^n
|\mathbf{x}\sim\mathcal{B}^{\frac{n}{2}}(\alpha)\}$ until
$|L^{(0)}_{2i-1}|\ge {n/2 \choose \alpha n/2}$.

\hskip .5cm
(1.2.2)
Repeat $L^{(0)}_{2i}\gets L^{(0)}_{2i}\cup\{(0^{n/2},\mathbf{x})\in\{0,1\}^n
|\mathbf{x}\sim\mathcal{B}^{\frac{n}{2}}(\alpha)\}$
until $|L^{(0)}_{2i}|\ge {n/2 \choose \alpha n/2}$.

\STATE Compute all level-$i$ lists $L^{(i)}_{j}$ for
$i=1,\dots,4,j=1,\dots,2^{4-i}$:

for $i = 1$ to $4$, do the following:

(2.1) Compute $k_i=\sum^{i-1}_{j=1}l_j$.

(2.2) Choose random $s^{(i)}_{j}\in\mathbb{Z}_{2^{k_i+l_i}}$
satisfying $s^{(i)}_{j}=s^{(i-1)}_{2j-1}+s^{(i-1)}_{2j}\mod 2^{k_i}$,
for all $j=1,\dots,2^{4-i}-1$.

(2.3) Compute $s^{(i)}_{2^{4-i}}=s-\sum^{2^{4-i}-1}_{j=1}s^{(i)}_{j}$.

(2.4) For $j = 1$ to $2^{4-i}$, do

\hskip .5cm
(2.4.1) Compute $L^{(i)}_{j}=filter(L^{(i-1)}_{2j-1}) \bowtie_{k_i+l_i}
filter(L^{(i-1)}_{2j})$.

\hskip .5cm
(2.4.2) Compute $filter(L^{(i)}_{j})$.

\STATE If $\exists\mathbf{e}\in L^{(4)}_{1}$ such that
$|\mathbf{e}|=\frac{n}{2}$, then return $\mathbf{e}$,
else return $\perp$.
\end{algorithmic}
\end{breakablealgorithm}

\begin{lemma} \cite{em19} \label{le:2.7}
The run time for sampling the level-$0$ lists in Algorithm \ref{al1} is $\widetilde{O}(2^{H(\alpha)n/2})$.
\end{lemma}

\vskip .2cm
{\bf Heuristic $1$ \cite{em19}.}
\vskip .2cm

Heuristically assume that the random variable that counts the number of
representations per run of the $EM^{(d)},d\ge 3$ algorithm is sharply
centered around its expectation to conclude that a single run
(or at most polynomially many runs) suffices to find a solution with good probability.
\vskip .2cm

This treatment is similar to that in Wagner’s original
$k$-tree algorithm \cite{w02} and its applications \cite{bcj11,hgj10}.

\begin{lemma}\cite{em19} \label{le:2.8}
Denote
\begin{eqnarray}
{\bf EMC1:}
\ \ \
7l_1+3l_2+l_3 &\le& (8H(\alpha)+\frac{1}{2}\log (8\alpha(1-\alpha)^{15}))n,
\label{E:2}
\\
{\bf EMC2:}
\ \ \
4l_1&\ge& (8H(\alpha)-4H(Bin_{2,\alpha}))n,
\label{E:3}
\\
{\bf EMC3:}
\ \ \
2l_2&\ge& (8H(\alpha)-2H(Bin_{4,\alpha}))n.
\label{E:4}
\end{eqnarray}
Then
\bi
\item
constraint {\bf EMC1} and Heuristic $1$ guarantees
that in a single run (or at most polynomially many runs) of
$EM^{(4)}$, the expected number of returned representations of the solution is at least one.

\item
Constraints {\bf EMC2}, {\bf EMC3} are necessary to
ensure that Heuristic 1 does not fail.
\ei
\end{lemma}

\begin{theorem}\cite{em19} \label{th:2.9}
Under Heuristic $1$, $EM^{(4)}$ solves the random subset sum problem in
time and memory $\widetilde{O}(2^{0.266n})$;
$EM^{(13)}$ reduces the time and memory complexity to
$\widetilde{O}(2^{0.255n})$.
\end{theorem}

\section{The Quantum Algorithm} \label{sec:3}

Now we use $EM^{(4)}$ as a bridge connecting random subset sum
problem and graph search problem. Recall the tree structure of
$EM^{(4)}$ in Figure \ref{fig1}. Denote by $L^{(i)}_{j}$ the
$j$-th list of the level $i$ lists in $EM^{(4)}$, where
$1\le i\le 4,1\le j\le 2^{4-i}$.

Consider the graph
\be
G_{search}(V_{search},E_{search}):=J(|L^{(0)}_{1}|,r)\times
J(|L^{(0)}_{2}|,r)\times\dots\times J(|L^{(0)}_{16}|,r),
\ee
which is the cartesian product of Johnson graphs.
\bi
\item
The vertices of $G_{search}$ are
$(U^{(0)}_{1},U^{(0)}_{2},\dots,U^{(0)}_{16})$, with
$U^{(0)}_{j}\subseteq L^{(0)}_{j},|U^{(0)}_{j}|=r,1\le j\le 16$.

\item
Denote by $U^{(i)}_{j}$ the $j$-th list of the level $i$ lists
that constructed from $U^{(0)}_{1}, U^{(0)}_{2}, \ldots, U^{(0)}_{16}$
according to $EM^{(4)}$, where $1\le i\le 4,1\le j\le 2^{4-i}$.

\item
For a vertex $(U^{(0)}_{1},U^{(0)}_{2},\ldots,U^{(0)}_{16})\in V_{search}$,
its data structure contains all $U^{(i)}_{j}$,
where $1\le i\le 4,1\le j\le 2^{4-i}$.

\item
A vertex $(U^{(0)}_{1},U^{(0)}_{2},\ldots,U^{(0)}_{16})\in V_{search}$
belongs to the marked set if and only if $U^{(4)}_{1}$ contains a solution to the
original random subset sum instance.
\ei
When we use quantum walk
to find a marked vertex, then based on the data structure of the marked vertex,
we can solve the original random subset sum problem.

In order to implement the quantum walk, it is necessary to
build the Johnson graphs $J(|L^{(0)}_{j}|,r)$ for $1\le j\le 16$.
So before we go on a walk, we need to build $L^{(0)}_{j}$ for
$1\le j\le 16$ by a classical sample, which is the same as the first step
of the $EM^{(4)}$ algorithm.

From now on, let
\be
L^{(i)}:=\mathbb{E}[|L^{(i)}_{j}|],\ \ \
U^{(i)}:=\mathbb{E}[|U^{(i)}_{j}|]
\ee
be the expected size of a list on level $i$ (before filtering)
for $0\le i\le 4$. Let
\be
L^{(i)}_f:=\mathbb{E}[|filter(L^{(i)}_{j})|],\ \ \
U^{(i)}_f:=\mathbb{E}[|filter(U^{(i)}_{j})|]
\ee
denote the expected size of filtered lists for $0\le i\le 4$.

According to Lemma \ref{le:2.8}, under Heuristic $1$, and
the constraints (\ref{E:2}),(\ref{E:3}) and (\ref{E:4}), if
$r =U^{(0)}= L^{(0)}$, then $M_{search}=V_{search}$.
That is, if $U^{(0)}_{j}$ contains all vertices of $L^{(0)}_{j}$
for $1\le j\le 16$, then all vertices of $G_{search}$ are marked.
Thus, $(\frac{U^{(0)}}{L^{(0)}})^{16}$ is the lower bound on the
probability that a vertex chosen randomly of $G_{search}$ is marked.

Let
\be
\epsilon=(\frac{U^{(0)}}{L^{(0)}})^{16}.
\ee
By
Lemma \ref{le:2.5},
$\delta=\Omega(1/U^{(0)})$. The remaining task is to determine $T_s,T_c,T_u$.

\vskip .2cm
{\bf Data Structure.}
\vskip .2cm

We use augmented radix trees \cite{bjlm13} to store the data structure of vertices in $V_{search}$. Augmented radix trees allow the three operations search, insertion and deletion in time logarithmic in the number of stored elements. Since our lists have exponential size and we ignore polynomials in the run time analysis, the cost of search, insertion and deletion operation can be ignored.

Recall that $EM^{(4)}$ level-$0$ lists are of the form $L^{(0)}_{j}=\{(\mathbf{c}^{(0)}_{j},\langle\mathbf{a},\mathbf{c}^{(0)}_{j}\rangle)\}$ for $1\le j\le 16$. For our $U^{(0)}_{j}\subseteq L^{(0)}_{j}$ we store the $\mathbf{c}^{(0)}_{j}$ and their inner products with $\mathbf{a}$ separately in $A^{(0)}_{j}=\{\mathbf{c}^{(0)}_{j}|\mathbf{c}^{(0)}_{j}\in U^{(0)}_{j}\}$ and $B^{(0)}_{j}=\{(\langle\mathbf{a},\mathbf{c}^{(0)}_{j}\rangle,\mathbf{c}^{(0)}_{j})|\mathbf{c}^{(0)}_{j}\in U^{(0)}_{j}\}$, where in $B^{(0)}_{j}$ elements are addressed via their first datum $\langle\mathbf{a},\mathbf{c}^{(0)}_{j}\rangle$, $1\le j\le 16$. Similarly, for $U^{(i)}_{j},1\le i\le 3,1\le j\le 2^{4-i}$, we also build separate $A^{(i)}_{j}$ and $B^{(i)}_{j}$. For $U^{(4)}_{1}$, it suffices to build $A^{(4)}_{1}$. So, we store $61$ sets in augmented radix trees.

Before computing $T_s,T_c,T_u$, we recall a classical list join operator that we will use in the analysis of complexity. The join operator performs the following task: given two lists of numbers $L_1$ and $L_2$ of respective sizes $|L_1|$ and $|L_2|$, together with two integers $M$ and $R$, the algorithm computes the list $L_0$ such that: $L_0=\{x_1+x_2|(x_1,x_2)\in L_1 \times L_2\land x_1+x_2\equiv R \mod  M\}$. The list $L_0$ can be constructed as follows. Sort $L_1$ and then for every $x_2\in L_2$ we find via binary search all elements $x_1\in L_1$ such that $x_1+x_2\equiv R \mod  M$. The complexity of this method is $\widetilde{O}(max (|L_1|,|L_2|,|L_0|))$\cite{w02}. Moreover, assuming that the values of the initial lists modulo $M$ are randomly distributed, the expected of $|L_0|$ is $\frac{|L_1|\cdot|L_2|}{M}$\cite{bcj11}.

Now, we compute $T_s,T_c,T_u$. Our goal is to obtain the optimal time complexity of our quantum algorithm, and then, under the optimal time complexity, to compute the memory complexity. We don't consider space-time tradeoffs. Thus, $T_s,T_c,T_u$ will be computed below are time complexity.

\vskip .2cm
{\bf Setup.}
\vskip .2cm

Start with analyzing the run time for sampling the level-$0$ lists in Algorithm \ref{al1}. Note that sampling is stopped when ${n/2 \choose \alpha n/2}=\widetilde{O}(2^{H(\alpha)n/2})$ different list elements have been found. Conclude by Lemma \ref{le:2.7} that this has only $\widetilde{O}(2^{H(\alpha)n/2})$ time complexity. So, $L^{(0)}=\widetilde{O}(2^{H(\alpha)n/2})$

Definition $\beta$ is the parameter that satisfies $U^{(0)}=(L^{(0)})^{\beta}$. Then, $U^{(0)}=\widetilde{O}(2^{\beta H(\alpha)n/2})$.

Now turn to the computation of the level-$1$ to level-$4$ lists.

Denote by $\gamma_i$ the probability
that a level-$i$ element gets filtered, that is $L^{(i)}_f=\gamma_i L^{(i)}$. Note that $L^{(1)}_f=L^{(1)}$, since level-$1$ elements are by construction in $\{0, 1\}^n$. For completeness, also define $L^{(0)}_f=L^{(0)}$. So, $\gamma_0=\gamma_1=1$. Let $l_4=n-l_1-l_2-l_3$.

Compute $\gamma_i$ by definition. The result is
\begin{displaymath}
\gamma_i=(1-\frac{2^{2}(i-1)\alpha^2}{((2^{i-1}-2)\alpha+2)^2})^nfor\;2\le i\le 3.
 \end{displaymath}

$U^{(i)}_{j}$ is constructed in the list join manner as above. The result is
\begin{equation}\label{E:5}
U^{(i)}=\frac{(U^{(i-1)}_f)^2}{2^{l_i}}=\frac{(\gamma_{i-1}U^{(i-1)})^2}{2^{l_i}}, \hbox{ for }
1\le i\le 4.
\end{equation}

Now solve for the $U^{(i)}$ from equation (\ref{E:5}). The result is
\be\ba{lll}
U^{(1)} &=& \widetilde{O}(2^{\beta H(\alpha)n-l_1}),\\

U^{(2)} &=& \widetilde{O}(2^{2\beta H(\alpha)n-2l_1-l_2}),\\

U^{(3)} &=& \widetilde{O}(2^{4\beta H(\alpha)n-4l_1-2l_2-l_3+2\log \gamma_2}),\\

U^{(4)} &=& \widetilde{O}(2^{8\beta H(\alpha)n-n-7l_1-3l_2-l_3+4\log \gamma_2+2\log \gamma_3}).
\ea\ee
Thus, the expected setup time complexity is
\begin{displaymath}
\mathbb{E}[T_s]=\max(U^{(0)}, \ U^{(1)}, \ U^{(2)}, \ U^{(3)}, \ U^{(4)}).
 \end{displaymath}

\begin{lemma}\cite{em19} \label{le:3.1}
Under Heuristic 1, $|U^{(i)}_j|=\widetilde{O}(U^{(i)})$ is
true for any $0\le i\le 4$ and $1\le j\le 2^{4-i}$.
\end{lemma}

\vskip .2cm
Proof.
Denote by $R^{(i)}_j$ the list made up of all representations in $U^{(i)}_j$,
where $0\le i\le 4, 1\le j\le 2^{4-i}$ and denote
$R^{(i)}:=\mathbb{E}[|R^{(i)}_{j}|]$. By Heuristic $1$,
$|R^{(i)}_j|=\widetilde{O}(R^{(i)})$.

The elements in the leaf lists $U^{(0)}_{j}\subseteq L^{(0)}_{j}$
are sampled from $\mathcal{B}^{\frac{n}{2}}(\alpha)\times 0^{\frac{n}{2}}$
and $0^{\frac{n}{2}}\times \mathcal{B}^{\frac{n}{2}}(\alpha)$.
As a consequence, the elements of the level-$1$ lists $U^{(1)}_{j}$ are from
$\mathcal{B}^{n}(\alpha)$.

Let $\mathbf{x}=\mathbf{c}^{(1)}_{1}+\dots+\mathbf{c}^{(1)}_{8}$, where
$\mathbf{c}^{(1)}_{j}\in U^{(1)}_{j},1\le j\le 8$. Then for each coordinate
$x_i$ of $\mathbf{x}$, $Pr[x_i=0]=(1-\alpha)^8$ and $Pr[x_i=1]=8\alpha(1-\alpha)^7$.
Hence a candidate $(\mathbf{c}^{(1)}_{1},\dots,\mathbf{c}^{(1)}_{8})
\in U^{(1)}_{1}\times\dots\times U^{(1)}_{8}$ is a representation of
the $n/2$-weight solution $\mathbf{e}$ with probability
\begin{displaymath}
p:=Pr[x_i=0]^{\frac{n}{2}}Pr[x_i=1]^{\frac{n}{2}}=(8\alpha(1-\alpha)^{15})^{\frac{n}{2}}.
 \end{displaymath}

So $|R^{(1)}_j|=|U^{(1)}_j|p,1\le j\le 8$ and $R^{(1)}=U^{(1)}p$.
Similarly, $|R^{(i)}_j|=|U^{(i)}_j|p, 2\le i\le 4, 1\le j\le 8$ and $R^{(1)}=U^{(1)}p$.
Furthermore, $|U^{(0)}_j|=\widetilde{O}(U^{(0)})=\widetilde{O}(2^{\beta H(\alpha)n/2}),
1\le j\le 16$.
So $|U^{(i)}_j|=\widetilde{O}(U^{(i)})$ for $0\le i\le 4$ and $1\le j\le 2^{4-i}$.

This finishes the proof.
\vskip .2cm

By Lemma \ref{le:3.1}, the setup time complexity is
\begin{displaymath}
T_s=\widetilde{O}(max(U^{(0)},U^{(1)},U^{(2)},U^{(3)},U^{(4)})).
 \end{displaymath}

{\bf Checking and Update}

Checking whether a vertex $(U^{(0)}_{1},U^{(0)}_{2},\dots,U^{(0)}_{16})$ is marked can be done easily by looking at $U^{(4)}_{1}$.
And the data of $U^{(4)}_{1}$ is already stored in $A^{(4)}_{1}$ after the setup subroutine. Thus, $T_c=\widetilde{O}(1)$.

One step of our random walk replaces a list item in exactly one of the leaf lists. We can perform one update by first deleting the replaced item and update the path to the root accordingly, and second adding the new item and again updating the path to the root.

We denote the operators that are used in the update subroutine as follows. $Insert(A^{(i)}_{j},\mathbf{x})$ inserts $\mathbf{x}$ into $A^{(i)}_{j}$, and $Delete(A^{(i)}_{j},\mathbf{x})$ deletes $\mathbf{x}$ from $A^{(i)}_{j}$. Furthermore, $\{\mathbf{x}\}\gets Search(B^{(i)}_{j},\langle\mathbf{a},\mathbf{y}\rangle)$ returns the list of all second datum $\mathbf{c}^{(i)}_{j}$ with first datum $\langle\mathbf{a},\mathbf{c}^{(i)}_{j}\rangle=\langle\mathbf{a},\mathbf{y}\rangle$, where $(\langle\mathbf{a},\mathbf{c}^{(i)}_{j}\rangle,\mathbf{c}^{(i)}_{j})\in B^{(i)}_{j}$.

Now, we describe the deleting an element subprogram. Without loss of generality, we assume the deleted element $\mathbf{x}\in U^{(0)}_{1}$.

\bu
\item $Delete(A^{(0)}_{1},\mathbf{x})$.

\item  $\{\mathbf{x}^{(0)}\}\gets
Search(B^{(0)}_{2},s^{(0)}_{1}-\langle\mathbf{a},\mathbf{x}\rangle\mod 2^{l_1})$.

\item For all $\mathbf{x}_{1}=\mathbf{x}+\mathbf{x}^{'}$ with
$\mathbf{x}^{'}\in \{\mathbf{x}^{(0)}\}$, do the following:
\bu
\item  $Delete(A^{(1)}_{1},\mathbf{x}_1)$.

\item $\{\mathbf{x}^{(1)}\}\gets Search(B^{(1)}_{2},s^{(1)}_{1}-\langle\mathbf{a},
\mathbf{x}_1\rangle\mod 2^{l_1+l_2})$.

\item  For all $\mathbf{x}_{2}=\mathbf{x}_1+\mathbf{x}^{'}$ with
$\mathbf{x}^{'}\in \{\mathbf{x}^{(1)}\}$, do the following:
\bu
\item $Delete(A^{(2)}_{1},\mathbf{x}_2)$.

\item $\{\mathbf{x}^{(2)}\}\gets Search(B^{(2)}_{2},s^{(2)}_{1}-\langle\mathbf{a},
\mathbf{x}_2\rangle\mod 2^{l_1+l_2+l_3})$.

\item For all $\mathbf{x}_{3}=\mathbf{x}_2+\mathbf{x}^{'}$ with
$\mathbf{x}^{'}\in \{\mathbf{x}^{(2)}\}$, do the following:
\bu

\item $Delete(A^{(3)}_{1},\mathbf{x}_3)$.

\item $\{\mathbf{x}^{(3)}\}\gets Search(B^{(3)}_{2},s^{(3)}_{1}
-\langle\mathbf{a},\mathbf{x}_3\rangle\mod 2^{n})$.

\item For all $\mathbf{x}_{4}=\mathbf{x}_3+\mathbf{x}^{'}$ with
$\mathbf{x}^{'}\in \{\mathbf{x}^{(3)}\}$, $Delete(A^{(4)}_{1},\mathbf{x}_4)$.
\eu
\eu
\eu
\eu

Since
\be
\mathbb{E}[|\{\mathbf{x}^{(0)}\}|]
=\frac{U^{(0)}}{2^{l_1}},\hskip .4cm
\mathbb{E}[|\{\mathbf{x}^{(1)}\}|]
=\frac{U^{(1)}}{2^{l_2}},\hskip .4cm
\mathbb{E}[|\{\mathbf{x}^{(2)}\}|]
=\frac{U^{(2)}}{2^{l_3}},\hskip .4cm
\mathbb{E}[|\{\mathbf{x}^{(3)}\}|]
=\frac{U^{(3)}}{2^{n-l_1-l_2-l_3}},
\ee
the expected cost of deleting an element is

\begin{displaymath}
\max(1,\frac{U^{(0)}}{2^{l_1}},\frac{U^{(0)}U^{(1)}}{2^{l_1+l_2}},\frac{U^{(0)}U^{(1)}U^{(2)}}{2^{l_1+l_2+l_3}},\frac{U^{(0)}U^{(1)}U^{(2)}U^{(3)}}{2^{n}}).
\end{displaymath}

Inserting an element is analogous to deleting an element.
Simply replace the deletion operator in the deleting subprogram with the insertion
operator to enable insertion of an element.
Thus, the expected update time complexity is
\begin{displaymath}
\mathbb{E}[T_u]=\max(1,\frac{U^{(0)}}{2^{l_1}},\frac{U^{(0)}U^{(1)}}{2^{l_1+l_2}},\frac{U^{(0)}U^{(1)}U^{(2)}}{2^{l_1+l_2+l_3}},\frac{U^{(0)}U^{(1)}U^{(2)}U^{(3)}}{2^{n}}).
\end{displaymath}

From Lemma \ref{le:3.1}, the update time complexity is
\begin{displaymath}
T_u=\widetilde{O}\left(\max\left(1,\  \frac{U^{(0)}}{2^{l_1}},\ \ \
\frac{U^{(0)}U^{(1)}}{2^{l_1+l_2}},\ \ \
\frac{U^{(0)}U^{(1)}U^{(2)}}{2^{l_1+l_2+l_3}},\ \ \
\frac{U^{(0)}U^{(1)}U^{(2)}U^{(3)}}{2^{n}}\right)\right).
\end{displaymath}

\vskip .2cm
{\bf Stopping unusually long updates.}
\vskip .2cm

The update time complexity is determined by the maximum cost over all vertices in a superposition. Therefore, even one node with an unusually slow update time complexity can disrupt our runtime. To prevent this problem, we modify our quantum walk algorithm by imposing an upper bound of $\kappa = poly(n)T_u$ steps for updating the data structure. After $\kappa$ steps, we simply stop the update of all nodes and proceed as if the update has been completed. We will use the following hypothesis.

\vskip .2cm
{\bf Heuristic $2$ \cite{hm18}.}
\vskip .2cm

Let $\epsilon$ be the fraction of marked states and $\delta$ be the spectral gap of the quantum walk $W$. Denote by $W_{stop}$ the quantum walk that forces $W$ to stop after $\kappa$ steps. Then the fraction of marked states in $W_{stop}$ is at least $\epsilon_{stop}=\widetilde{\Omega}(\epsilon)$, and
the spectral gap of $W_{stop}$ is at least $\delta_{stop}=\widetilde{\Omega}(\delta)$.
Moreover, the stationary distribution of $W_{stop}$ is close to the distribution of its setup.
Namely, we obtain with high probability a random node which can be superposition of the Johnson graph with correctly built data structure.

In summary, we use $EM^{(4)}$ as a bridge connecting random subset sum problem and graph search problem. Then, we give a new algorithm, which starts with a classic sample, next implements a quantum walk over the cartesian product of Johnson graphs. By reasonably constructing the data structure, we calculate the $T_s,T_c,T_u$.
From Theorem \ref{th:2.2} and Lemma \ref{le:2.8}, we know that under Heuristic $1$, Heuristic $2$ and constraints (\ref{E:2}), (\ref{E:3}) and (\ref{E:4}) there exists an algorithm that with high probability finds a marked vertex in time
\begin{equation}\label{E:6}
T=L^{(0)}+T_s+\frac{1}{\sqrt{\epsilon}}(\frac{1}{\sqrt{\delta}}T_u+T_c).
\end{equation}
,where the time complexity of classic sampling is $L^{(0)}$ and the time complexity of quantum walk is $T_s+\frac{1}{\sqrt{\epsilon}}(\frac{1}{\sqrt{\delta}}T_u+T_c)$.

We will give the optimal value of all parameters later. Now let us give our quantum $EM^{(4)}$ algorithm.

\begin{breakablealgorithm}\label{al2}
\caption{Quantum $EM^{(4)}$ Algorithm.}
\begin{algorithmic}[1]
{\bf Input:} subset sum instance $(\mathbf{a},s)\in(\mathbb{Z}_{2^{n}})^{n+1}$;

\vskip .2cm
parameters $\alpha,\beta\in[0,1],l_1,l_2,l_3\in\mathbb{N}$ and $l_1+l_2+l_3\le n$.

\vskip .2cm
{\bf Output:} solution $\mathbf{e}\in\{0,1\}^{n}$ with $|\mathbf{e}|=\frac{n}{2}$, or
$\perp$ if no solution is found.

\vskip .2cm
\STATE
[Classical Sampling]Sample all level-$0$ lists $L^{(0)}_{i}$ for $i=1,\dots,16$:

\vskip .2cm
\STATE
[Quantum Setup] Prepare the initial state:
\be
|\pi\rangle=\frac{1}{c}\left(\sum_{\stackrel{
(U^{(0)}_{1}\times
\dots\times U^{(0)}_{16})}
{\in V_{search}}}\
\bigotimes^{16}_{j=1}\ \
|U^{(0)}_{j}\rangle\, |coin\rangle\, |data\rangle\right),
\ee
where $c$ is the normalization factor, $|coin\rangle$ is
the superposition of vertices adjacent to the current vertex,
and $|data\rangle$ is the data structure of the current vertex.

Note that parameters $\alpha,\beta,l_1,l_2,l_3$ are used in the process
of building $|data\rangle$.

\vskip .2cm
\STATE Do the following $O(1/\sqrt{\epsilon})$ times:

\vskip .2cm
(3.1) [Quantum Checking]
\begin{displaymath}
 \bigotimes^{16}_{j=1}\ |U^{(0)}_{j}\rangle\,
 |coin\rangle\, |data\rangle\mapsto
 \left\{
\begin{array}{ll}
-\bigotimes^{16}_{j=1}\
|U^{(0)}_{j}\rangle\,|coin\rangle\,|data\rangle, & \hbox{ if }
 (U^{(0)}_{1}\times\dots\times U^{(0)}_{16})\in M, \\
\\

\bigotimes^{16}_{j=1}\ |U^{(0)}_{j}\rangle\,|coin\rangle\,|data\rangle,\
& \hbox{ else}.
                   \end{array}
                 \right.
\end{displaymath}

\vskip .2cm
(3.2) [Quantum Update] Do the following $O(1/\sqrt{\delta})$ times:

\vskip .2cm
\hskip .5cm
(3.2.1) Take a quantum step of the walk.

\vskip .2cm
\hskip .5cm
(3.2.2) Update the data structure accordingly, stop after $\kappa$ steps.

\vskip .2cm
\STATE Measure the final state, obtain a state
$\bigotimes^{16}_{j=1}|U^{(0)}_{j}\rangle|coin\rangle|data\rangle$.

If $\exists \mathbf{e}\in A^{(4)}_{1}$ such that $|\mathbf{e}|=\frac{n}{2}$,
then return $\mathbf{e}$, else return $\perp$.

\end{algorithmic}
\end{breakablealgorithm}

\begin{theorem} \label{th:4.1}
Under Heuristics 1,2 and the constraints (\ref{E:2}), (\ref{E:3}) and (\ref{E:4}),
Algorithm \ref{al2} gives with high probability the solutions
of random subset sum instances
in time and with memory $2^{0.209n}$.
\end{theorem}

\vskip .2cm
Proof.
By the analysis of section \ref{sec:3}, the time complexity of Algorithm \ref{al2} is
\be
T=L^{(0)}+T_s+\frac{1}{\sqrt{\epsilon}}(\frac{1}{\sqrt{\delta}}T_u+T_c),
\ee
where
\be\ba{lll}
L^{(0)} &=& \widetilde{O}(2^{H(\alpha)n/2}),\\

U^{(0)} &=& \widetilde{O}(2^{\beta H(\alpha)n/2}),\\

\epsilon &=& \ds \left(\frac{U^{(0)}}{L^{(0)}}\right)^{16},\\

\delta &=& \ds \Omega(1/U^{(0)}),\\

T_c &=& \widetilde{O}(1),\\

T_s &=& \widetilde{O}(\max(U^{(0)},U^{(1)},U^{(2)},U^{(3)},U^{(4)})),\\

T_u &=& \ds \widetilde{O}(\max(1,\ \ \
\frac{U^{(0)}}{2^{l_1}},\ \ \
\frac{U^{(0)}U^{(1)}}
{2^{l_1+l_2}},\ \ \
\frac{U^{(0)}U^{(1)}U^{(2)}}{2^{l_1+l_2+l_3}},\ \ \
\frac{U^{(0)}U^{(1)}U^{(2)}U^{(3)}}{2^{n}})).
\ea\ee

Furthermore,
\be\ba{lll}
U^{(1)}&=& \ds \widetilde{O}(2^{\beta H(\alpha)n-l_1}),\\

U^{(2)}&=& \ds \widetilde{O}(2^{2\beta H(\alpha)n-2l_1-l_2}),\\

U^{(3)}&=& \ds \widetilde{O}(2^{4\beta H(\alpha)n-4l_1-2l_2-l_3+2\log \gamma_2}),\\

U^{(4)}&=& \ds \widetilde{O}(2^{8\beta H(\alpha)n-n-7l_1-3l_2-l_3+4
\log \gamma_2+2\log \gamma_3}),\\

\gamma_0 &=& 1,\\

\gamma_i&=& \ds (1-\frac{2^{2}(i-1)\alpha^2}{((2^{i-1}-2)\alpha+2)^2})^n, \hbox{ for }
i=1,2,3.
\ea\ee
So
\be\ba{r}
\ds
T\ =\ \widetilde{O}\left(\max\left((U^{(0)})^{\frac{1}{\beta}},
\frac{(U^{(0)})^2}{2^{l_1}},
\frac{(U^{(0)})^4}{2^{2l_1+l_2}},
\frac{\gamma^2_2(U^{(0)})^8}{2^{4l_1+2l_2+l_3}},
\frac{\gamma^2_3\gamma^4_2(U^{(0)})^{16}}{2^{n+7l_1+3l_2+l_3}},\right.\right.
\\
\\

\ds\left.\left.
\frac{(U^{(0)})^{\frac{8}{\beta}-6.5}}{2^{l_1}},
\frac{(U^{(0)})^{\frac{8}{\beta}-4.5}}{2^{2l_1+l_2}},
\frac{(U^{(0)})^{\frac{8}{\beta}-0.5}}{2^{4l_1+2l_2+l_3}},
\frac{\gamma^2_2(U^{(0)})^{\frac{8}{\beta}+7.5}}{2^{n+7l_1+3l_2+l_3}}\right)\right).
\ea
\ee

Under the constraints (\ref{E:2}), (\ref{E:3}) and (\ref{E:4}), the numerical
optimization for minimizing $T$ gives
\begin{displaymath}
\alpha=0.188,\ \ \
\beta=0.941,\ \ \
l_1=0.184n,\ \ \
l_2=0.209n,\ \ \
l_3=0.188n.
\end{displaymath}
So $T=\widetilde{O}(2^{0.209n})$ and $L^{(0)}=\widetilde{O}(2^{0.209n})$.

Obviously, $L^{(0)}\le C_{memory} \le T$,
where $C_{memory}$ represents the memory cost.
Thus, under Heuristic 1, and constraints (\ref{E:2}), (\ref{E:3}), (\ref{E:4}),
Algorithm \ref{al2} runs in time $T=2^{0.209n}$ using $L^{(0)}=2^{0.209n}$ memory.

This finishes the proof.
\vskip .2cm

\begin{remark}
In a similar way to quantum $EM^{(4)}$ algorithm, we can quantize
$EM^{(d)}$ algorithm, for $d\ge 3$. When analyzing varying depths,
we could not improve over the run time. For $3\le d\le 8$, our
results are listed in Table $1$. If $\beta=1$, then our
quantum $EM^{(d)}$ algorithm is the classic $EM^{(d)}$ algorithm,
for $d\ge 3$. As observed from the Table $1$, $\beta$ is
getting closer to $1$ as the depth increases. We conjecture that
$|T_{quantum EM}(d)-T_{EM}(d)|$ converges for $d \to \infty$,
where $T_{quantum EM}(d)$ is the run time of our quantum $EM^{(d)}$
algorithm and $T_{EM}(d)$ is the run time of $EM^{(d)}$ algorithm.
\end{remark}

\begin{center}
{\small
    \begin{center}
    \centerline{\small {\bf Table 1}~~Comparison of the run times of quantum and classical $EM$ algorithm by depth $d$.}\vskip 1mm
    \label{Tab:1}

    {\small
    \begin{tabular*}{8.5cm}{ccccc}
        &   & \ \ \ \ \;\;\;\; \;\; quantum EM &   & \hskip 1.5cm EM     \\[0.8ex]
        \cmidrule(r){3-4}\cmidrule(r){5-5}
                        & $d$   & $T$ & $\beta$ & $ \hskip 1.5cm T$    \\[0.8ex]
        \midrule
            &3   &$2^{0.2531}$  &$0.8889$    &$ \hskip 1.5cm 2^{0.2960}$   \\[0.8ex]
            &4   &$2^{0.2090}$  &$0.9412$    &$ \hskip 1.5cm 2^{0.2659}$  \\[0.8ex]
            &5   &$2^{0.2194}$  &$0.9697$    &$ \hskip 1.5cm 2^{0.2616}$ \\[0.8ex]
            &6   &$2^{0.2326}$  &$0.9846$    &$ \hskip 1.5cm 2^{0.2584}$ \\[0.8ex]
            &7   &$2^{0.2417}$  &$0.9922$    &$ \hskip 1.5cm 2^{0.2565}$ \\[0.8ex]
            &8   &$2^{0.2473}$  &$0.9961$    &$ \hskip 1.5cm 2^{0.2558}$ \\[0.8ex]
        \bottomrule
    \end{tabular*}}
\end{center}}
\end{center}


\acknowledgements{\rm This work is supported partially by the National Natural
Science Foundation of China under Grant No.
11671388, and CAS Project QYZDJ-SSW-SYS022.}



\begin{thebibliography}{199}


\bibitem{a98}
Ajtai M. The shortest vector problem in $L_2$ is np-hard for randomized reductions.
In: {\it Proc. 30th annual ACM symposium on Theory of computing}, pp. 10-19. ACM, 1998.

\bibitem{a07}
Ambainis A. Quantum walk algorithm for element distinctness. {\it SIAM J. Computing}
 37(1): 210-239, 2007.

\bibitem{aakv01}
Aharonov D, Ambainis A, Kempe J, and Vazirani U. Quantum walks on graphs. In:
{\it Proc. 33rd annual ACM symposium on Theory of computing}, pp. 50-59. ACM, 2001.

\bibitem{b84}
Brickell E F. Solving low density knapsacks. In: {\it Advances in Cryptology}, pp. 25-37.
Springer, 1984.

\bibitem{bcj11}
Becker A, Coron J, and Joux A. Improved generic algorithms for
 hard knapsacks. In: {\it Annual International Conference on the Theory
 and Applications of Cryptographic Techniques}, pp. 364-385. Springer, 2011.

\bibitem{bjlm13}
Bernstein D J, Jeffery S, Lange T, and Meurer A. Quantum algorithms for the subset-sum
problem. In: {\it International Workshop on Post-Quantum Cryptography}, pp. 16-33.
Springer, 2013.

\bibitem{clos91}
Coster M J, LaMacchia B A, Odlyzko A M, and Schnorr C P.
An improved low-density subset sum algorithm. In: {\it Workshop on the Theory and
Application of of Cryptographic Techniques}, pp. 54-67. Springer, 1991.

\bibitem{cr84}
Chor B, Rivest R L. A Knapsack type public key cryptosystem based on arithmetic in finite
fields. In: Blakley GR, Chaum D (eds.) {\it Advances in Cryptology - CRYPTO’84}
(Aug 19-23, 1984).
Lecture Notes in Computer Science 196,
pp. 54-65. Springer, Heidelberg, Germany; Santa Barbara, CA, USA.

\bibitem{em19}
Esser A and May A. Better sample - Random subset sum in $2^{0.255n}$ and its
Impact on decoding random linear codes. arXiv:1907.042951v1.

\bibitem{fmv16}
Faust S, Masny D, Venturi D. Chosen-ciphertext security from subset sum. In:
Cheng CM, Chung KM,
Persiano G, Yang BY (eds.) {\it PKC 2016: 19th International Conference on Theory
and Practice of Public Key
Cryptography, Part I} (Mar 6-9, 2016).
Lecture Notes in Computer Science 9614, pp. 35-46.
Springer, Heidelberg, Germany;
Taipei, Taiwan.

\bibitem{g96}
Grover L K. A fast quantum mechanical algorithm for database search. In:
{\it Proc. 28th annual ACM symposium on Theory of computing}, pp. 212-219. ACM, 1996.

\bibitem{gj79}
Garey M R and Johnson D S. {\it Computers and Intractability: A Guide to the Theory of NP-
Completeness}. W H Freeman, 1979.

\bibitem{gm91}
Galil Z and Margalit O. An almost linear-time algorithm for the dense subset-sum  problem.
{\it SIAM J Computing} 20(6): 1157-1189, 1991.

\bibitem{hgj10}
Howgrave-Graham N and Joux A. New generic algorithms for hard knapsacks. In: {\it
 Annual International Conference on the Theory and Applications of Cryptographic
 Techniques}, pp. 235-256. Springer, 2010.

\bibitem{hm18}
Helm A and May A. Subset Sum Quantumly in $1.17^n$. In: {\it
13th Conference on the Theory of Quantum Computation, Communication and Cryptography} (TQC 2018),
Article No. 5, pp. 5:1-5:16.


\bibitem{hs74}
Horowitz E and Sahni S. Computing partitions with applications to the knapsack problem.
{\it Journal of the ACM} (JACM) 21(2): 277-292, 1974.

\bibitem{in96}
Impagliazzo R and Naor M. Efficient cryptographic schemes provably as secure as subset sum.
{\it Journal of Cryptology}
9(4): 199-216, 1996.

\bibitem{js91}
Joux A and Stern J. Improving the critical density of the lagarias-odlyzko
 attack against subset sum problems. In: {\it International Symposium on Fundamentals of
 Computation Theory}, pp. 258-264. Springer, 1991.

\bibitem{kt17}
Kachigar G and Tillich J P. Quantum information set decoding algorithms.
504 CoRR, abs/1703.00263, 2017.

\bibitem{lo85}
Lagarias J C and Odlyzko A M. Solving low-density subset sum problems.
{\it Journal of the ACM} (JACM) 32(1): 229-246, 1985.

\bibitem{lps10}
Lyubashevsky V, Palacio A and Segev G. Public-key cryptographic primitives provably as
secure as subset sum.
In: Micciancio D (ed.) {\it TCC 2010: 7th Theory of Cryptography Conference}
(Feb 9-11, 2010).
Lecture Notes in Computer Science 5978, pp. 382-400. Springer, Heidelberg, Germany;
Zurich, Switzerland.

\bibitem{mh78}
Merkle R, Hellman M. Hiding information and signatures in trapdoor knapsacks.
{\it IEEE T. Information
Theory} 24(5): 525-530, 1978.

\bibitem{mnrs11}
Magniez F, Nayak A, Roland J, and Santha M. Search via quantum
walk. {\it SIAM J. Computing}, 40(1): 142-164, 2011.


\bibitem{s04}
Szegedy M. Quantum speed-up of Markov chain based algorithms. In: {\it Proc.
45th IEEE Symposium on Foundations of Computer Science}, pp. 32-41,
2004.

\bibitem{s08}
Santha M. Quantum walk based search algorithms. In: {\it Proc. 5th TAMC}, pp. 31-46.
arXiv:0808.0059, 2008.

\bibitem{sll19}
Shao C,  Li Y, and  Li H. Quantum algorithm design:
Techniques and applications. {\it J. Systems Science and Complexity} 32: 375-452, 2019.

\bibitem{ss81}
Schroeppel R and  Shamir A. A $t=O(2^{n/2})$, $s=O(2^{n/4})$ algorithm
for certain np-complete problems. {\it SIAM J. Computing}, 10(3): 456-464, 1981.

\bibitem{w02}
Wagner D. A generalized birthday problem. In: Yung, M. (ed.)
{\it Advances in Cryptology - CRYPTO 2002}.
Lecture Notes in Computer Science, vol. 2442, pp. 288-303.
Springer, Heidelberg, Germany, Santa Barbara, CA,
USA, Aug 18-22, 2002.

























\end{thebibliography}
\end{document}